\documentclass{elsart}

\usepackage{graphicx}
\usepackage{amsmath}

\begin{document}

\begin{frontmatter}
\bibliographystyle{elsart-num}

\title{Entanglement sudden death in qubit-qutrit systems}

\author[Ann]{Kevin Ann}
\ead{kevinann@bu.edu} and
\author[Jaeger]{Gregg Jaeger}
\ead{jaeger@bu.edu, correspondence author}

\address[Ann]{Department of Physics, Boston University \\
590 Commonwealth Avenue, Boston, MA 02215}
\address[Jaeger]{Quantum Imaging Lab, Department of Electrical and Computer Engineering,\\
and Division of Natural Sciences, Boston University, Boston, MA 02215}

\date{\today}

\begin{abstract}
We demonstrate the existence of entanglement sudden death (ESD), the
complete loss of entanglement in finite time, in qubit-qutrit
systems. In particular, ESD is shown to occur in such systems
initially prepared in a one-parameter class of entangled mixed
states and then subjected to local dephasing noise. Together with
previous results, this proves the existence of ESD for some states
in \emph{all} quantum systems for which rigorously defined
mixed-state entanglement measures have been identified. We
conjecture that ESD exists in all quantum systems prepared in
appropriate bipartite states.
\end{abstract}

\begin{keyword}
decoherence,
disentanglement,
entanglement sudden death,
mixed-state entanglement,
phase noise
\PACS 03.65.Yz, 03.67.-a, 03.67.Mn, 42.50.Lc
\end{keyword}

\end{frontmatter}

\section{Introduction}
An important pursuit common to both the investigation of quantum
information processing and quantum measurement is fully
understanding the behavior of entangled systems in their
environments. This is so because entanglement both enables important
quantum protocols and arises in the quantum measurement, both of
which in practice take place in a larger environment that affects
their fundamental properties. When studying quantum information
processing, it is almost always necessary to take into account the
inevitable interaction of the involved entangled systems with their
environment. The resulting influence can in many cases be described
simply as local dephasing noise, which tends to decohere the quantum
subsystems and to affect system entanglement. Even subject only to
such simply weak noise, suitable error correction or mitigation
methods must typically be implemented for processing to be reliable.
An understanding of decoherence and disentanglement is essential for
the assessment of the necessity and scope of such techniques, which
tend to be costly in resource terms. Because both entanglement and
decoherence are similarly significant in modeling quantum
measurement, their study also contributes to our understanding of
the measurement process.

The influence of environmental noise on entangled quantum systems
differs in general from that on both classical systems and separable
quantum systems. For example, recent work by Dodd and Halliwell
\cite{DH} and Yu and Eberly \cite{YE03,YE04,YE06a,YE06b,YE06c,YE07}
has invalidated the intuition that the effects of weak local noises
on composite quantum systems are necessarily additive in general, as
it is in the latter two cases. The former authors showed that all
bipartite initially Gaussian states of two-particle systems
described by continuous valued dynamical variables disentangle in
finite times under local decoherence effects only, exponentially
reducing composite-system coherence. The latter authors showed that
although the influences of weak noises on coherence in the simplest
bipartite quantum system, the two-qubit system, are additive, the
influence of noise on \emph{entanglement} is not additive. They also
found, similarly to the latter duo, that a combination of local weak
noise influences on two-qubit systems can lead, for some state
classes, to the striking phenomenon of the total loss of
entanglement in finite time, which they termed {\it entanglement
sudden death} (ESD) \cite{YE06a,YESCI}, rather exhibiting
exponential entanglement decay as the coherence so decays. Moreover,
this phenomenon has recently been experimentally confirmed in an
all-optical experimental realization of a two-qubit system in a
local dephasing noise environment \cite{ADH07}. Here, we show that
ESD in finite-dimensional systems is not limited only to two-qubit
systems but can occur in composite systems of larger dimension as
well.

Here, we demonstrate the existence of ESD in the hybrid qubit-qutrit
system, by considering a class of states in which the qubit and
qutrit are individually incoherent but may still possess a degree of
\emph{composite-system} coherence and entanglement. The two-qubit
and the qubit-qutrit system are the only finite-dimensional
bipartite systems for which rigorously defined general mixed-state
entanglement measures are known; the bipartite Gaussian states are
the only infinite-dimensional such systems \cite{VW02,Jaeger06}.
Given the above results, this result supports our conjecture that
entanglement sudden death is a generic phenomenon, in the sense of
extending to all bipartite quantum systems when appropriately
prepared, because it establishes the occurrence of ESD in all
quantum systems in which a rigorous general analysis is possible,
given the limited scope of mixed-state entanglement measures
currently available.

The remainder of this paper is organized as follows. In Section 2,
we motivate the choice of initial qubit-qutrit state class
considered and provide explicit descriptions of local and
multi-local dephasing noise in the operator-sum decomposition. In
Section 3, we demonstrate entanglement sudden death for this class
of states in environments producing either type of dephasing noise.
A summary of conclusions and our ESD conjecture are discussed in
Section 4.

\section{Initial States and Noise Model}

We begin by motivating the choice of class of bipartite mixed states
and of the dephasing noise model here used to demonstrate the
existence of the phenomenon of entanglement sudden death in
qubit-qutrit systems. This class includes entangled states with
sufficient susceptibility to local dephasing noise that ESD always
occurs in any system initially prepared in them when subject to it.

\subsection{Class of qubit-qutrit states of interest}
The general density matrix describing the composite qubit-qutrit
system is $\rho_{\rm AB}=[\rho_{ij}]$, where
$\rho^*_{ji}=\rho_{ij},\ \sum_i\rho_{ii}=1,$ with $i,j=1,\ldots,6$.
There are two sorts of term in this matrix: that responsible for
subsystem coherence and that responsible for joint-system coherence,
both in a chosen basis. The latter terms are typically associated
with quantum entanglement. It is important to recognize, however,
that entanglement, by definition, is basis-independent whereas
coherence is not. It is therefore also noteworthy that
basis-dependent noise can dramatically influence entanglement. Let
us designate the first and second subsystems by `A' and `B',
respectively. Each of the individual subsystem states, described by
a reduced density matrix, is obtained by tracing over the variables
associated with the other subsystem.

The general reduced density matrix $\rho_{\rm A}$ for the qubit is
thus
\begin{equation}
\rho_{\rm A} = \left(
\begin{array}{cc}
 (\rho_{11} + \rho_{22} + \rho_{33}) & (\rho_{14} + \rho_{25} + \rho_{36}) \\
 (\rho_{41} + \rho_{52} + \rho_{63}) & (\rho_{44} + \rho_{55} + \rho_{66})
\end{array}\right).
\end{equation}
States in which the subsystems are incoherent can be ones in which
the composite system nonetheless possesses high joint-state
coherence and is entangled. We therefore begin by taking the reduced
states obtained from $\rho_{\rm AB}$ to be incoherent in this
representation to more directly observe the influence of local
dephasing noise on global properties, entanglement in particular. We
do so by taking all off-diagonal terms in the reduced density
matrices to be zero \cite{Cun07}. Thus, we seek joint-system states
consistent with the incoherent reduced-state
$$\rho'_{\rm A} ={\rm diag}\big( (\rho_{11} + \rho_{22} +
\rho_{33}),(\rho_{44} + \rho_{55} + \rho_{66})\big)\ .\nonumber$$
Similarly, because the qutrit reduced density matrix $\rho_{\rm B}$
for the qutrit are
\begin{equation}
\rho_{\rm B} = \left(
\begin{array}{cccccc}
 (\rho_{11} + \rho_{44}) & (\rho_{12} + \rho_{45}) & (\rho_{13} + \rho_{46})  \\
 (\rho_{21} + \rho_{54}) & (\rho_{22} + \rho_{55}) & (\rho_{23} + \rho_{56}) \\
 (\rho_{31} + \rho_{64}) & (\rho_{32} + \rho_{65}) & (\rho_{33} + \rho_{66})
\end{array}\right)\ ,
\end{equation}
we consider the incoherent qutrit reduced state
$$\rho'_{\rm B} ={\rm diag}\big((\rho_{11} + \rho_{44}),(\rho_{22} + \rho_{55}),(\rho_{33} +
\rho_{66})\big)\ .\nonumber$$ The composite-system density matrices
that yield reduced states of the forms $\rho'_{\rm A}$ and
$\rho'_{\rm B}$ is
\begin{equation}
\hspace{8pt}\rho'_{\rm AB} = \left(
\begin{array}{cccccc}
 \rho_{11} & 0 & 0 & 0 & \rho_{15} & \rho_{16} \\
 0 & \rho_{22} & 0 & \rho_{24} & 0 & \rho_{26} \\
 0 & 0 & \rho_{33} & \rho_{34} & \rho_{35} & 0 \\
 0 & \rho_{42} & \rho_{43} & \rho_{44} & 0 & 0 \\
 \rho_{51} & 0 & \rho_{53} & 0 & \rho_{55} & 0 \\
 \rho_{61} & \rho_{62} & 0 & 0 & 0 & \rho_{66}
\end{array}\right) ,
\label{generalAnsatz}
\end{equation}
where $\rho_{ij}\in\Re$; the zeroed off-diagonal terms must each be
so, given that they have been chosen to be nonnegative and real,
because for each of them there is an off-diagonal term in either
$\rho'_{\rm A}$ or $\rho'_{\rm B}$ that is a sum  containing it
({\it cf.} Eqs. 1-2) that is zero. The nonzero terms in the full
joint-system density matrix of Eq. 3 are exclusively those related
to joint-system global state coherence and typically also
entanglement. Nonetheless, as we see below, they are affected by
local noise.

For all mixed states $\rho_{\rm AB}$ of qubit-qutrit systems,
entanglement is well quantified by the negativity,
$\mathcal{N}(\rho_{\rm AB})$, which is the degree to which a
positive map fails to be completely positive
\cite{VW02,HHH96,Per96,Hor97}. For our class of states, this is
given by the absolute value of the negative eigenvalues
$\lambda^{\rm T_A(-)}_{k}$ of the partial transpose of the
full-system density matrix $\rho'_{\rm AB}$ with respect to the
smaller dimensional subsystem, when at least one exists, and by zero
otherwise. Thus, here,
\begin{equation}
\mathcal{N}(\rho'_{\rm AB}) ={\rm max}\big\{0,\sum_{k} |\lambda^{\rm
T_A(-)}_{k}|\big\}\ .
\end{equation}

\vfill\eject

For simplicity and definiteness, let us study the specific
one-parameter class, within those with density matrices described by
Eq. 3, of the form
\begin{equation}
\hspace{8pt}\rho'_{AB}(\bar{x})= \left(
\begin{array}{cccccc}
 \ \ \frac{1}{4} \ \ & \ \ 0 \ \ & \ \ 0 \ \ & \ \ 0 \ \ & \ \ 0 \ \ & \ \ \bar{x} \ \ \\
 0 & \frac{1}{8} & 0 & 0 & 0 & 0 \\
 0 & 0 & \frac{1}{8} & 0 & 0 & 0 \\
 0 & 0 & 0 & \frac{1}{8} & 0 & 0 \\
 0 & 0 & 0 & 0 & \frac{1}{8} & 0 \\
 \bar{x} & 0 & 0 & 0 & 0 & \frac{1}{4}
\end{array}
\label{ansatzWithZ} \right)\ ,
\end{equation}
where $0\le \bar{x} \le \frac{1}{4}$. This choice is made because
taking the diagonal terms to be not all equal prevents the density
matrix and its partial transpose from having identical eigenvalues.
This is a necessary condition for $\rho'_{\rm AB}(\bar{x})$ to be
separable for some physically allowed values of $\bar{x}$ into which
initially entangled states of this class evolve, under the
time-evolution corresponding noise model chosen below, because all
states remain within this class under that evolution, as shown
below. The range of values of $\bar{x}$ is necessary for the matrix
to be a well-defined density matrix, see below. ESD also takes place
for a larger class of states of the form of Eq. 3, as it does for
this class, for example, when the conditions (i)
$\rho_{24},\rho_{34}$ and $\rho_{35}$ are all zero, (ii) the
diagonal elements are as chosen here, and (iii) one takes
$\rho_{15}=\rho_{16}=\rho_{26}=\bar{x}$, are all satisfied. However,
in that case, the expressions for eigenvalues are more complex
without obvious added benefit for exhibiting or describing the
phenomenon. The matrices described by Eq. 5 are well-defined density
matrices because they satisfy the following necessary conditions for
any matrix $\rho$ to be a density matrix: (1) Hermiticity of $\rho$,
(2) unitarity of ${\rm tr}\rho$, and (3) positive semi-definiteness
of $\rho$. Conditions (1) and (2) are obviously satisfied by this
matrix. Condition (3) is satisfied because all eigenvalues of
$\rho'_{AB}(\bar{x})$ remain non-negative only within the above
range of values of $\bar{x}$, as will be shown below where these are
exhibited.

\subsection{Noise model}
In order to exhibit ESD in this class, it suffices to consider local
dephasing noise alone acting on subsystems that are dynamically
isolated from each another. The most general time-evolved
open-system density matrix expressible in the operator-sum
decomposition is
\begin{equation}
\rho\left(t\right) = \mathcal{E}\left(\rho\left(0\right)\right) =
\sum_{\mu}K_{\mu}\left(t\right)\rho\left(0\right)
K_{\mu}^{\dagger}\left(t\right) \label{krausGeneral}\ ,
\end{equation}
where the $K_\mu(t)$, which satisfy a completeness condition
guaranteeing that the evolution be trace-preserving, represent the
influence of statistical noise which can be global or local in
scope, and where the index runs over the number of elements required
for the decomposition.  For local and multi-local dephasing
environments, the $K_{\mu}(t)$ are of the form  $K_{\mu}(t)=
F_{j}(t)E_{i}(t)$, so that
\begin{eqnarray}
\rho_{\rm AB}\left(t\right) =
\mathcal{E}\left(\rho\left(0\right)\right) = \sum_{i = 1}^{2}
\sum_{j = 1}^{3} F_{j}\left(t\right)E_{i}\left(t\right) \rho_{\rm
AB}\left(0\right)
E_{i}^{\dagger}\left(t\right)F_{j}^{\dagger}\left(t\right)
\label{krausSpecific},
\end{eqnarray}
where
\begin{eqnarray}
E_{1}(t) &=& {\rm diag}(1,\gamma_{\rm A}) \otimes {\rm diag}(1,1,1)
=
{\rm diag}(1,1,1,\gamma_{\rm A},\gamma_{\rm A},\gamma_{\rm A})\ , \\
E_{2}(t) &=& {\rm diag}(0,\omega_{\rm A}) \otimes {\rm diag}(1,1,1)
=
{\rm diag}(0,0,0,\omega_{\rm A},\omega_{\rm A},\omega_{\rm A})\ , \\
F_{1}(t) &=& {\rm diag}(1,1) \otimes {\rm diag}(1,\gamma_{\rm
B},\gamma_{\rm B})=
{\rm diag}(1,\gamma_{\rm B},\gamma_{\rm B},1,\gamma_{\rm B},\gamma_{\rm B})\ , \\
F_{2}(t) &=& {\rm diag}(1,1) \otimes {\rm diag}(0,\omega_{\rm B},0)
=
{\rm diag}(0,\omega_{\rm B},0,0,\omega_{\rm B},0)\ , \\
F_{3}(t) &=& {\rm diag}(1,1) \otimes {\rm diag}(0,0,\omega_{\rm B})
= {\rm diag}(0,0,\omega_{\rm B},0,0,\omega_{\rm B})\ ,
\end{eqnarray}
$\gamma_{\rm A}\left(t\right) = e^{-t(\Gamma_{\rm A}/2)},
\gamma_{\rm B}\left(t\right) = e^{-t(\Gamma_{\rm A}/2)},\
\omega_{\rm A}\left(t\right) = \sqrt{1-\gamma_{{\rm A}}^{2}(t)},
\omega_{\rm B}\left(t\right) = \sqrt{1-\gamma_{{\rm B}}^{2}(t)}\\(X
= {\rm A}, {\rm B})$. The $E_{i}(t)$ and $F_{j}(t)$ induce local
dephasing in the qubit and qutrit states, respectively, and
individually satisfy the usual completeness condition for the
operator-sum decomposition of CPTP maps \cite{Jaeger06}, the
$\Gamma_X$ quantifying the rate of local exponential dephasing. The
noise parameters in Eqs. 10-12 have been chosen such that the rate
of dephasing from levels 1 and 2 relative to the state 0 are equal,
to simplify resulting expressions. The time-dependence of
$\gamma(t)$'s are also often left implicit in the sequel to lend
compactness to expressions, particularly in the explicit forms of
density matrices.

\section{Entanglement Sudden Death}
Here, we show that qubit-qutrit systems initially prepared in the
simple one-parameter family of states $\rho'_{\rm AB}(\bar{x})$
motivated and constructed above exhibit entanglement sudden death.
In order to find the entanglement of the qubit-qutrit system, we
first find the eigenvalues $\left\{ \lambda^{\rm T_A}_{k}(x,t)
\right\}$ ($k = 1, \ldots, 6$) of the partial transpose with respect
to the qubit of $\rho'_{AB}(x,t)$, and then sum over the absolute
values of the negative ones to obtain the negativity. The three
specific environmental noise situations now considered in turn are:
(1) qubit dephasing only, (2) qutrit dephasing only, and (3)
combined local dephasing.

\subsection{Qubit (A) dephasing noise only}
The time-dependent solution of Eq. \ref{krausSpecific}, with the
initial density matrix given in Eq. \ref{ansatzWithZ}, in the case
when the $E_{i}(t)$ are of the general form of Eqs. 8-9 and when no
noise from the environment of the qutrit is present, so that
$F_{i}(t)={\rm I}$, is
\begin{equation}
\hspace{4pt} \rho'_{\rm AB}(x,t) = \left(
\begin{array}{cccccc}
 \ \frac{1}{4} \  &  \ 0 \ & \ 0 \ & \ 0 \ & \ 0 \ & x\gamma_{\rm A} \  \\
 0 & \frac{1}{8} & 0 & 0 & 0 & 0 \\
 0 & 0 & \frac{1}{8} & 0 & 0 & 0 \\
 0 & 0 & 0 & \frac{1}{8} & 0 & 0 \\
 0 & 0 & 0 & 0 & \frac{1}{8} & 0 \\
 x \gamma_{\rm A} & 0 & 0 & 0 & 0 & \frac{1}{4}
\end{array}
\label{qubitADephasingOnlyZ}
\right).
\end{equation}
Its eigenvalues are $\left\{ \lambda_{k}(x,t) \right\}=\{
\frac{1}{8}, \frac{1}{8}, \frac{1}{8}, \frac{1}{8}, \frac{1}{4}(1 +
4 x \gamma_{\rm A}), \frac{1}{4}(1 - 4 x \gamma_{\rm A})\}$. This is
a well-defined density matrix as long as its eigenvalues are
non-negative. The only eigenvalue of this matrix that can possibly
become negative for positive values of $x$ is the last one, which
occurs when $\bar{x}=x\gamma_A > \frac{1}{4}$, which values lies
outside the range of values of $\bar{x}$ of our class, since
$\gamma_A(t)\leq 1$ for all times $t$ and $x$ is strictly less than
or equal that $\frac{1}{4}$, so it remains a well-defined density
matrix, as required and discussed above. Taking into account the
form of $\gamma_A(t)$, one sees that decoherence takes the form of
the exponentially decay of off-diagonal terms of the density matrix,
which terms tend to zero only in the large-time limit
$t\rightarrow\infty$: 
both nonzero off-diagonal elements of the matrix of Eq. 14 decay
asymptotically to zero as a function of time for all values of $x$,
in particular within the restricted range of physically allowed
values.

Because previous analyses have shown that disentanglement proceeds
at least as fast as decoherence in a broad range of states of
two-qubit and two-qutrit systems \cite{YE03,AJ07}, we expect this
result to hold for the hybrid qubit-qutrit system as well. This is
indeed the case. To see that this is so, consider the eigenvalues of
the partial transpose of $\rho'_{\rm AB}(x,t)$ with respect to qubit
A, namely, $\left\{ \lambda^{\rm T_A}_{k}(x,t) \right\}= \{
\frac{1}{4}, \frac{1}{4}, \frac{1}{8}, \frac{1}{8}, \frac{1}{8}(1 +
8 x \gamma_{\rm A}), \frac{1}{8}(1 - 8 x \gamma_{\rm A})\}$.
Clearly, the only one of these eigenvalues that can become negative
is the last one. Thus, in accord with Eq. 4, one finds the degree of
entanglement as a function of time to be
\begin{equation}
\mathcal{N}(\rho'_{\rm AB}(x,t))=\max\{0,x \gamma_{\rm A}(t)-1/8\}\
, \label{generalNegativitiesA}
\end{equation}
for the range of values of $x$ specifying states within the class of
states under consideration. Thus, one sees that all such states
which are initially entangled, that is for which $x>{1\over 8}$,
become separable for all values of $x\gamma_{\rm A}(t)\leq
\frac{1}{8}$, that is, irreversibly disentangled as soon as
$\gamma_{\rm A}(t)$ reaches ${1\over 8x}$. By contrast, decoherence
occurs only asymptotically in the large-time limit
$t\rightarrow\infty$, in which $\gamma_{\rm A}(t)\rightarrow 0$.

\subsection{Qutrit (B) dephasing noise only}
For noise acting on qutrit B alone, the time evolved density matrix
is similarly
\begin{equation}
\hspace{8pt}\rho'_{\rm AB}(x,t) = \left(
\begin{array}{cccccc}
 \ \frac{1}{4} \  &  \ 0 \ & \ 0 \ & \ 0 \ & \ 0 \ & x\gamma_{\rm B} \  \\
 0 & \frac{1}{8} & 0 & 0 & 0 & 0 \\
 0 & 0 & \frac{1}{8} & 0 & 0 & 0 \\
 0 & 0 & 0 & \frac{1}{8} & 0 & 0 \\
 0 & 0 & 0 & 0 & \frac{1}{8} & 0 \\
 x \gamma_{\rm B} & 0 & 0 & 0 & 0 & \frac{1}{4}
\end{array}
\label{qubitBDephasingOnlyZ}
\right).
\end{equation}
The eigenvalues $\left\{ \lambda^{\rm T_A}_{k}(x,t) \right\}$ of the
partial transpose of $\rho'_{\rm AB}(x,t)$ with respect to qubit A
are $ \{ \frac{1}{4}, \frac{1}{4}, \frac{1}{8}, \frac{1}{8},
\frac{1}{8}(1 + 8 x \gamma_{\rm B}), \frac{1}{8}(1 - 8 x \gamma_{\rm
B})\}$. As in the previous case, the only eigenvalue that can
potentially be negative is the last one. Thus, we similarly find the
degree of entanglement of the composite system to be
\begin{equation}
\mathcal{N}(\rho'_{\rm AB}(x,t))=\max\{0,x \gamma_{\rm B}(t)-1/8\}\
. \label{generalNegativitiesA}
\end{equation}
One sees that the state to irreversibly become separable once
$x\gamma_B \leq \frac{1}{8}$, that is, as soon as $\gamma_{\rm
B}(t)$ reaches ${1\over 8x}$. Entanglement sudden death therefore
also takes place in the case of local dephasing noise acting on the
qutrit alone. By contrast, and as in the case of qubit dephasing,
full decoherence of the composite system occurs only in the
large-time limit, $t\rightarrow\infty$.

\subsection{Multi-local dephasing noise}
Finally, as one might now expect given the above results for local
dephasing noise, multi-local dephasing noise affecting both
components of the system gives rise to the time evolved state
\begin{equation}
\hspace{8pt}\rho'_{\rm AB}(x,t) = \left(
\begin{array}{cccccc}
 \ \frac{1}{4} \  &  \ 0 \ & \ 0 \ & \ 0 \ & \ 0 \ & x\gamma_{\rm A} \gamma_{\rm B}  \\
 0 & \frac{1}{8} & 0 & 0 & 0 & 0 \\
 0 & 0 & \frac{1}{8} & 0 & 0 & 0 \\
 0 & 0 & 0 & \frac{1}{8} & 0 & 0 \\
 0 & 0 & 0 & 0 & \frac{1}{8} & 0 \\
 x \gamma_{\rm A} \gamma_{\rm B} & 0 & 0 & 0 & 0 & \frac{1}{4}
\end{array}
\label{qubitABDephasing}
\right).
\end{equation}
The timescales of decay of off-diagonal terms, that is, of
decoherence are simple products of terms appearing due to individual
local dephasing noise contributions shown in the previous two
subsections separately, so that decoherence is additive, a result
anticipated on the basis of previous study of two-qubit systems
under local noise \cite{AJ07}. One finds that the off-diagonal terms
decay at an even faster rate than in the previous cases of dephasing
noise on either qubit or qutrit subsystem alone but, again, the
coherence decays exponentially to zero only in the limit
$t\rightarrow\infty$. The eigenvalues $\left\{ \lambda^{\rm
T_A}_{k}(x,t) \right\}$ of the partial transpose with respect to
qubit A are $ \{ \frac{1}{4}, \frac{1}{4}, \frac{1}{8}, \frac{1}{8},
\frac{1}{8}(1 + 8 x \gamma_{\rm A}\gamma_{\rm B}), \frac{1}{8}(1 - 8
x \gamma_{\rm A}\gamma_{\rm B})\}$. As before, the only eigenvalue
that can be negative is the last one. Thus, we find
\begin{equation}
\mathcal{N}(\rho'_{\rm AB}(x,t))=\max\{0,x \gamma_{\rm
A}(t)\gamma_{\rm B}(t)-1/8\}\ , \label{generalNegativitiesA}
\end{equation}
so that, by the same argument as made in the previous cases,
multi-local dephasing noise is seen to induce \emph{entanglement
sudden death} for the chosen state class, and does so even more
quickly than in the single local dephasing noise cases considered
above, namely, the state of the qubit-qutrit system irreversibly
becomes separable as soon as $\gamma_{\rm A}(t)\gamma_{\rm
B}(t)\le\frac{1}{8x}$.

\section{Conclusion}
We have shown that entanglement sudden death (ESD) can take place in
hybrid qubit-qutrit systems, in particular, for a straightforwardly
motivated and simple one-parameter class of mixed states. This
result extends those of Dodd and Halliwell and of Yu/Eberly on ESD,
and supports the conjecture we have made here that ESD is a generic
phenomenon in the sense of existing in all bipartite quantum states,
by showing that it occurs also in the remaining case not considered
by them that is capable of general examination within the current
limitations of entanglement measure theory. In addition to quantum
information processing applications, this result has implications
for the decoherence program for addressing the measurement problem:
the abrupt loss of entanglement between subsystems may shed light on
how the environment picks out a preferred basis and how a quantum
system transitions towards classical behavior under the decoherence
model of quantum measurement.

The definitive demonstration of the phenomenon of entanglement
sudden death in all finite-dimensional bipartite quantum systems
remains an open problem, one which can only be definitely addressed
if one possesses an entanglement measure for mixed states of
qu-d-it--qu-d$'$-it systems for general values of $d,d'$ and a class
of states of such systems to which it would apply. We anticipate
demonstrating entanglement sudden death in such systems in a future
publication. Finally, we note that, in a very recent preprint,
Che\c{c}ki\'nska and W\'odkiewicz have studied entangled
qutrit-pairs in noisy atomic channels and found evidence that ESD
may occur in that situation, which would represent additional
progress toward this goal \cite{CW}.



\end{document}